\documentclass[preprint]{aastex}

\newcommand{\degree}{\mbox{$^{\circ}$}}
\newcommand{\ghz}{\mbox{GHz}}
\newcommand{\kms}{\mbox{km\,s$^{-1}$}}
\newcommand{\yr}{\mbox{yr$^{-1}$}}
\newcommand{\m}{\mbox{m}}

\newcommand{\methanol}{\mbox{CH$_3$OH}}
\newcommand{\oh}{\mbox{OH}}
\newcommand{\sio}{\mbox{SiO}}
\newcommand{\water}{\mbox{H$_2$O}}

\shortauthors{BAE ET AL.}
\shorttitle{A MULTI-EPOCH SIMULTANEOUS WATER AND METHANOL MASER SURVEY}

\begin{document}

\title{A Multi-Epoch, Simultaneous Water and Methanol Maser Survey Toward Intermediate-Mass Young Stellar Objects}

\author{Jae-Han Bae\altaffilmark{1}, Kee-Tae Kim\altaffilmark{1}, So-Young Youn\altaffilmark{1}, Won-Ju Kim\altaffilmark{1,2}, Do-Young Byun\altaffilmark{1}, \\ Hyunwoo Kang\altaffilmark{1,3}, and Chung Sik Oh\altaffilmark{1}}
\altaffiltext {1} {Korea Astronomy and Space Science Institute, Hwaam-Dong, Yuseong-Gu, Daejeon 305-348, Korea; whorujh@kasi.re.kr, ktkim@kasi.re.kr}
\altaffiltext {2} {Department of Astronomy and Space Science, Chungnam National University, Daejeon 305-764, Korea} 
\altaffiltext {3} {Department of Physics and Astronomy, FPRD, Seoul National University, Seoul 151-742, Korea} 

\begin{abstract}

We report a multi-epoch, simultaneous 22~\ghz\ \water\ and 44~\ghz\ Class~I \methanol\ maser line survey
toward 180 intermediate-mass young stellar objects,
including 14 Class~0, 19 Class~I objects, and 147 Herbig Ae/Be stars.
We detected \water\ and \methanol\ maser emission toward 16 (9~\%) and 10 (6~\%) sources 
with 1 new \water\ and 6 new \methanol\ maser sources. 
The detection rates of both masers rapidly decrease as the central (proto)stars evolve, 
which is contrary to the trends in high-mass star-forming regions.
This suggests that the excitations of the two masers are closely related to
the evolutionary stage of the central (proto)stars and
the circumstellar environments.
\water\ maser velocities deviate on average 9~\kms\ from the ambient gas velocities
whereas \methanol\ maser velocities match quite well with the ambient gas velocities. 
For both maser emissions, large velocity differences 
($\left| v_{\rm{H_{2}O}} - v_{\rm{sys}} \right| > 10~\kms$ 
and $\left| v_{\rm{CH_{3}OH}} - v_{\rm{sys}} \right| > 1~\kms$) are mostly confined to Class~0 objects.
The formation and disappearance of \water\ masers is frequent and 
their integrated intensities change by up to two orders of magnitude.
In contrast, \methanol\ maser lines usually show no significant change in
intensity, shape, or velocity.
This is consistent with the previous suggestion that
\water\ maser emission originates from the base of an outflow
while 44~\ghz\ Class~I \methanol\ maser emission arises from
the interaction region of the outflow with the ambient gas.
The isotropic maser luminosities are well correlated with the bolometric luminosities of the central objects.
The fitted relations are
$L_{\rm{H_{2}O}} = 1.71 \times 10^{-9} (L_{\rm{bol}})^{0.97}$
and $L_{\rm{CH_{3}OH}} = 1.71 \times 10^{-10} (L_{\rm{bol}})^{1.22}$.

\end{abstract}


\section{INTRODUCTION}

Maser emission is an important signpost of star formation, especially at the early stages.
Since the discovery of the first maser line, the \oh\ emission at a frequency of 1665~MHz by \citet{Weaver65}, 
many surveys have been performed toward star-forming regions at various frequencies, e.g., 
1.66540~\ghz\ and 1.66736~\ghz\ (\oh\ maser lines), 
6.66852~\ghz\ (Class II~\methanol\ $5_{1}-6_{0}~A^{+}$ maser line), 
22.23508~\ghz\ (\water\ $6_{16}-5_{23}$ maser line), 
42.82059~\ghz\ and 43.12208~\ghz\ (\sio\ $v = 1, 2, J = 1-0$ maser lines), 
and 44.06943~\ghz\ (Class~I \methanol\ $7_{0}-6_{1}~A^{+}$ maser line).

22~\ghz\ \water\ masers are thought to be pumped by excitation of the \water\ rotation states 
in collisions with $\rm{H_{2}}$ molecules \citep{Elitzur89}.
Previous surveys of \water\ masers have shown that 
this maser is frequently observed toward young stellar objects (YSOs) 
over a wide range of masses (\citealt{Genzel77, Rodriguez80})
and exhibits a wide range of relative velocities 
with respect to the parental dense cores (\citealt{Felli92, Furuya05, Breen10, Caswell10}).
High resolution studies of the individual YSOs have revealed that 
\water\ masers are located very close ($\lesssim 1000~\rm{AU}$) to the central 
(proto)stars (\citealt{Claussen98, Marti99, Seth02, Furuya05}).

\methanol\ masers are the most recent maser species to be intensively studied.
There are numerous transitions that can be divided into two classes, 
I and II, according to the classification of \citet{Menten91}.
Class~I \methanol\ maser transitions include the $4_{-1}-3_{0}~E$, $7_{0}-6_{1}~A^{+}$, $5_{-1}-4_{0}~E$, and $8_{0}-7_{1}~A^{+}$ lines 
at 36, 44, 84, and 95~\ghz, 
while Class~II \methanol\ maser transitions include the $5_{1}-6_{0}~A^{+}$, $2_{0}-3_{-1}~E$, 
and $J_{0}-J_{-1}~E$ lines at 6.7, 12, and 157~\ghz.
Even though both classes are commonly observed in (high-mass) star-forming regions,
their pumping mechanisms are known to be different.
Using theoretical models, \citet{Cragg92} suggested that Class~I \methanol\ masers result from collisional excitation, 
while class~II \methanol\ masers appear when there is a source of continuum radiation.

The \methanol\ maser at 44~\ghz\ is the most common and strongest Class~I maser.
Nevertheless, only $\sim200$ sources have been detected until now \citep{Valtts10}.
44~\ghz\ Class~I \methanol\ masers are thought to be associated mainly with high-mass star formation, 
yet interferometric observations found that they are usually offset by $\sim0.1-1$~pc from other signposts of star formation, 
such as ultracompact HII (UCHII) regions, infrared sources, and \water\ masers (\citealt{Kurtz04, Cyganowski09}).
The observed 44~\ghz\ \methanol\ maser velocities are different from the ambient molecular gas velocities by 
$\lesssim$~10~\kms\ (\citealt{Bachiller90, Fontani10}), 
which is much smaller compared to \water\ maser relative velocities.

Intermediate-mass stars have masses of 2$-$10~$M_{\odot}$. 
These lower and upper boundaries are of importance.
First, pre-main sequence (PMS) stars with masses $>$~$2~M_{\odot}$ are expected to follow fully radiative tracks once the quasi-static contraction has ended \citep{Waters98}.
This implies that these stars could evolve in different ways from lower-mass stars.
Second, stars more massive than $\sim10~M_{\odot}$ spend their whole PMS stage as obscured objects \citep{Waters98}, 
which means that they are not visible until they reach the main sequence.
Intermediate-mass YSOs (IMYSOs) share many characteristics with their high-mass counterparts, 
but many of them are located far closer to the Sun and are less deeply embedded \citep{Alonso09}. 
Thus, it would be helpful to study IMYSOs to understand high-mass star formation.
 
Despite the importance of studying IMYSOs,
no systematic maser line surveys toward them have been undertaken so far.
Maser observations toward IMYSOs were usually made as a part of surveys targeting low-mass YSOs
(e.g., \citealt{Claussen96, Furuya03}).
Otherwise, maser surveys have been limited to Herbig Ae/Be (HAeBe) stars
(e.g., \citealt{Palla93}),
which are recognized as intermediate-mass PMS stars 
(see \citealt{The94} and references therein):
have spectral types of A or earlier, with emission lines; 
lie in an obscured region; 
and illuminate fairly bright nebulosity in their immediate vicinity.

In this study, we present a multi-epoch, simultaneous \water\ and 44~\ghz\ Class~I \methanol\ maser line survey 
toward 180 IMYSOs in various evolutionary stages.
This paper is organized as follows.
The source selection and the observations are described in Section 2 and 3, respectively.
We present the results with some comments on the individual sources in Section 4 and 
discuss the observational results in Section 5.
We summarize the main results in Section 6.

\section{SOURCE SELECTION}

To define our sample of IMYSOs, 
we collected sources from published studies satisfying one of the following two criteria: 
(1) protostellar or PMS objects with bolometric luminosities of $40~L_{\odot} < L_{\rm{bol}} \lesssim 1000~L_{\odot}$, 
which are thought to be the precursors of intermediate-mass stars; 
or (2) known or candidate HAeBe stars.
All selected sources have declinations greater than $-30\degree$.  

We first selected objects that satisfy criterion (1):
14 Class~I objects from \citet{Saraceno96}; 
4 Class~0, 3 Class~0/I, 5 Class~I, 1 HAeBe, and 9 unclassified sources from \citet{Furuya03}; 
4 Class~0, 5 Class~0/I, 4 Class~I, and 1 unclassified source from \citet{Froebrich05}; 
7 Class~0 and 2 Class~I objects from \citet{Alonso10}.
To clarify the evolutionary stages of the collected unclassified objects, 
we consulted additional classification works, which are
noted in Column 5 of Table~\ref{table:source}.
Considering the overlapped sources from the references mentioned above, 
we produced a sample containing 14 Class~0, 19 Class~I, and 2 HAeBe objects from the first criterion.

We then collected objects satisfying criterion (2).
They are primarily selected from catalogs published by \citet{The94} and \citet{Vieira03}.
\citet{The94} presented 108 Be and Ae stars, 
which were historically recognized as true members or potential candidates of the HAeBe stellar group,
and \citet{Vieira03} suggested 131 HAeBe stars and candidates.
Of members in the two catalogs, 
86 sources in \citet{The94} and 76 sources in \citet{Vieira03} are north of $-30$\degree\ with 22 sources in common.
Our sample also contains five other HAeBe stars:
BD+24\degree676, BD+41\degree3731, V517~Cyg, and V1057~Cyg from \citet{Palla93}, and
V1318~CygS from \citet{Palla95}.
Thus, the complete sample satisfying the second criterion would be 145 HAeBe stars.

In summary, our sample consists of 180 IMYSOs: 14 Class~0, 19 Class~I, and 147 HAeBe stars.
The objects in our sample are listed in Table~\ref{table:source}.

\section{OBSERVATIONS}

We carried out a multi-epoch, simultaneous survey of
\water\ $6_{16} - 5_{23}$ (22.23508~\ghz)
and class~I \methanol\ $7_{0}-6_{1} A^{+}$ (44.06943~\ghz)
maser lines toward 180 IMYSOs.
The observations were performed
using the Korean Very Long Baseline Interferometry Network (KVN) $21~\m$ radio telescope
at the Yonsei station over three different periods.
The first observations were conducted toward all of the 180 IMYSO samples during 21 days from 2010 January to May.
On 2010 October 20 and December 2, the second and third epochs were observed toward the 26 objects 
that showed maser emission in the first epoch.

The telescope is equipped with a multi-frequency receiving system,
which makes it possible to observe at both frequencies simultaneously \citep{Han08}.
The 86~\ghz\ and 129~\ghz\ receivers will be added to the receiving system
in 2011 for simultaneous observations in four different frequency bands. 
We used 4096 channel Digital Spectrometers, each with 32~MHz bandwidth. 
Table~\ref{table:obs} presents a summary of the observations.

The pointing and focus checks were made every 2 hr
using nearby known strong \water\ maser sources.
The pointing accuracy was maintained to better than 5$''$ during the observations.
The data were calibrated by the standard chopper wheel method
and the line intensity was obtained on the $T_{A}^{*}$ scale.
The conversion factors between $T_{A}^{*}$ and the flux density
are 11.1~Jy~K$^{-1}$ at 22~\ghz\ and 11.6~Jy~K$^{-1}$ at 44~\ghz.

All the observations were made using position switching mode
with total (ON+OFF) integration time of 30 minutes,
which typically yields about $0.5~\textrm{Jy}$ rms noise levels
for both maser transitions at $0.21~\kms$ velocity resolution.
Since the KVN telescopes are of the shaped Cassegrain type,
they have a smaller full width at half-maximum (FWHM) primary beam and
much higher ($\sim$14~dB) first sidelobe levels than
conventional Cassegrain antennas with the same size primary reflectors
(\citealt{Kim11}; Lee et al. 2011, in preparation).
The first sidelobes are separated by about 1.5 times the FWHM from the
primary beam center.
To determine whether the detected maser emission is contaminated
by nearby bright maser sources, we mapped an area of
1.5 FWHM $\times$ 1.5 FWHM around each source with detectable maser
emission using half beam spacing.
All data were reduced with the CLASS software.

\section{RESULTS}

\subsection{Overall Results}

\begin{figure*}
 \centering
 \includegraphics{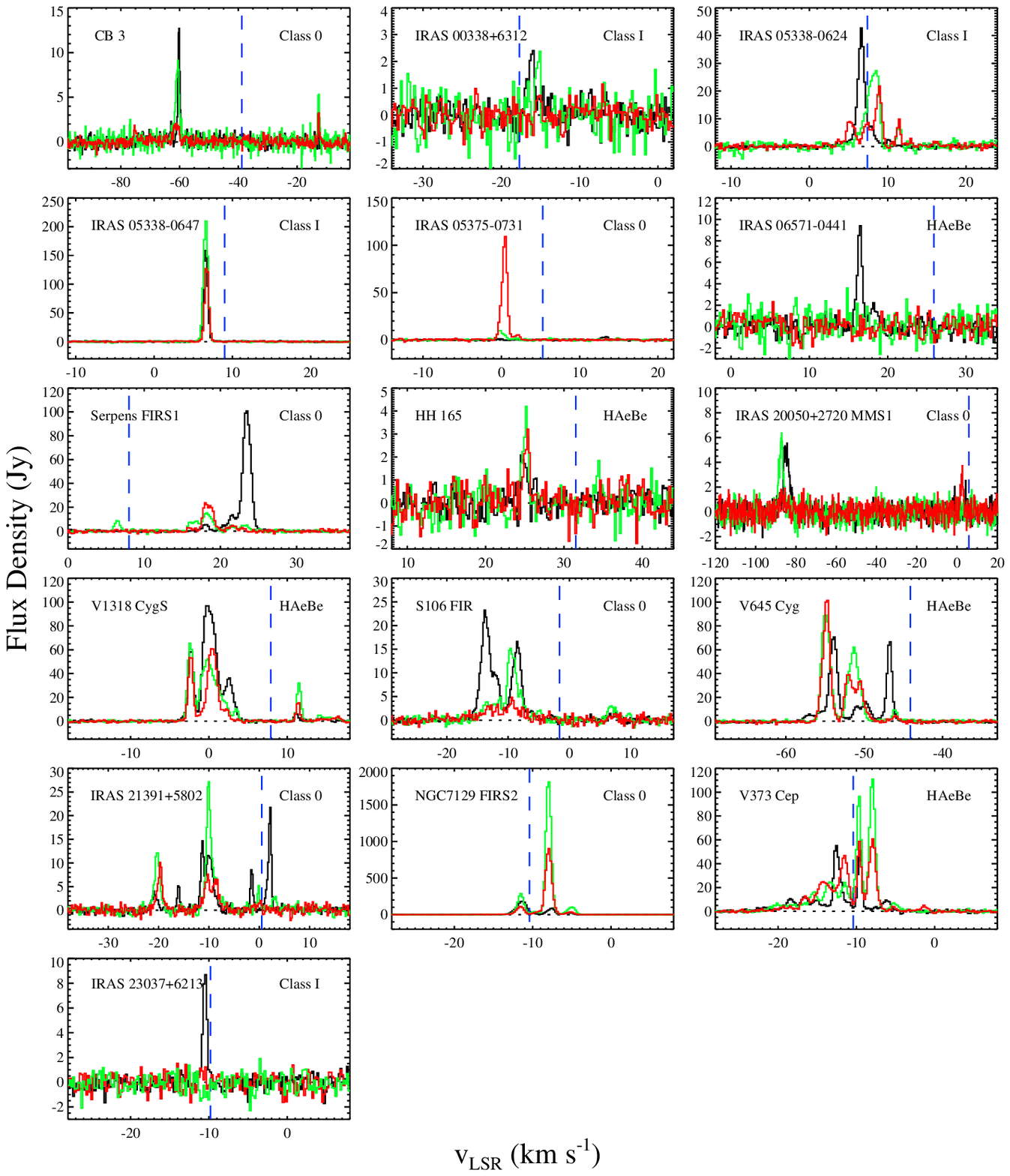}
 \caption{Spectra of the \water\ maser sources. 
In each panel, black, green, and red lines represent data obtained in the first, second, and third observation periods, respectively.
Blue vertical dashed line indicates the systemic velocity of the parental dense molecular core.
The evolutionary stage of the central object is shown at the upper right corner.\label{fig:water}}
\end{figure*}

\begin{figure*}
 \centering
 \includegraphics{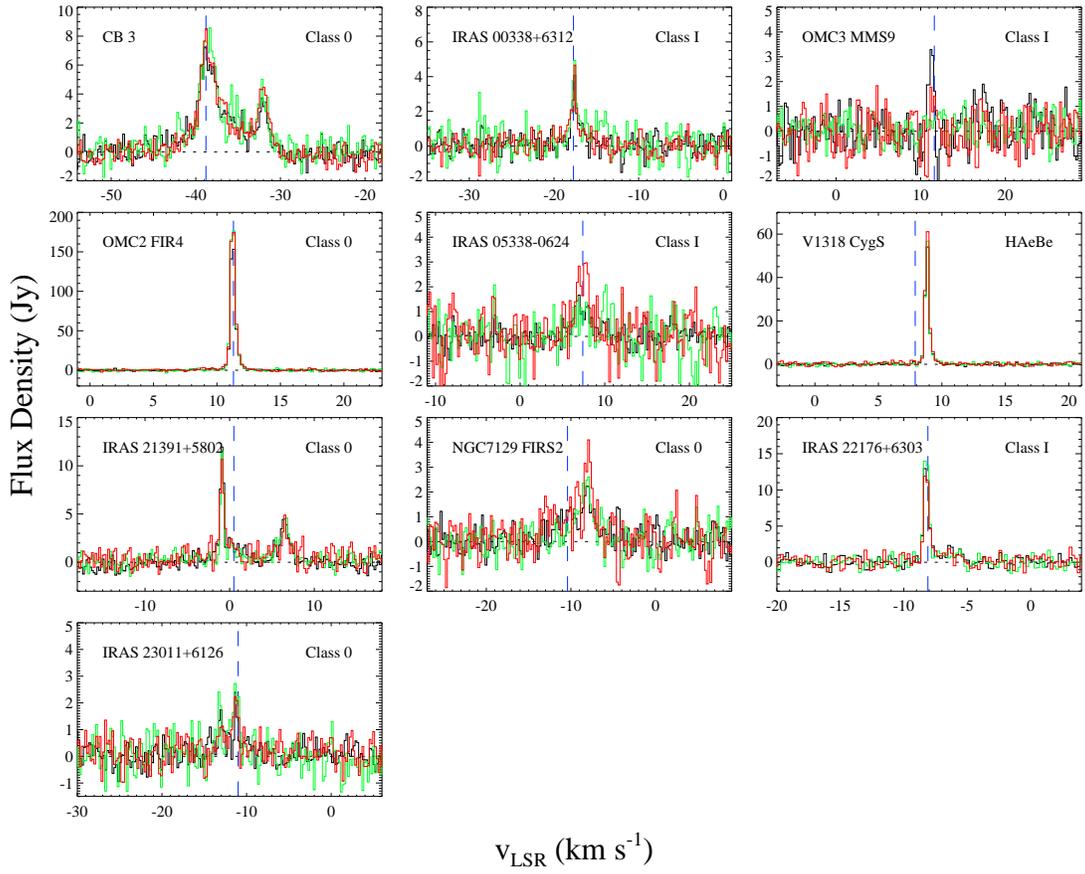} 
 \caption{Same as in Figure~\ref{fig:water} but for spectra of the 44~\ghz\ Class~I \methanol\ maser sources.\label{fig:methanol}}
\end{figure*}

Of the 180 IMYSOs observed, we detected \water\ and \methanol\ 
maser emission toward 20 and 12 sources, respectively (Figures~\ref{fig:water} and \ref{fig:methanol}).
All the detected maser lines are limited to signals stronger than the 3$\sigma$ rms noise level.
However, several objects are located close enough to one another that they fall within a single beam or its sidelobes.
Further analysis shows that four \water\ and two \methanol\ masers do not arise from the target IMYSOs.
Those sources will be discussed in detail below.
In total, 16 sources show \water\ maser emission and 10 sources show \methanol\ maser emission.
Six objects emit both \water\ and \methanol\ maser emission.
Consequently, the detection rates of \water\ masers and \methanol\ masers are $9~\%$ (16/180) and $6~\%$ (10/180), respectively. 
One intriguing source is IRAS 20050+2720 MMS1, which shows extremely 
blueshifted ($\sim-90~\kms$) \water\ maser line emission with respect to the molecular gas.
\citet{Furuya03} suggested that this emission might be related to
an extremely high-velocity CO outflow emanating from the central
protostar \citep{Bachiller95}.
The observed properties of all detected \water\ and \methanol\ maser lines
are presented in Tables \ref{table:P22} and \ref{table:P44}:
the source name in Column 1, the observing date in Column 2,
the Gaussian-fitted line flux, peak velocity, and line width (FWHM) in Columns 3 - 5, 
the integrated line flux, peak intensity, and peak velocity in Columns 6 - 8,
and the equivalent width ($\int{F_{\nu}~dv}$/$F_{\rm peak}$) in Column 9.

Table \ref{table:detection} exhibits the detection rates of \water\ and \methanol\ masers 
for the sources in different evolutionary stages.
The detection rates of both masers decrease substantially as the central (proto)stars evolve.
The detection rate of \water\ masers is $50~\%$ for Class~0 objects, 
while the rates are $21~\%$ and $3~\%$ for Class~I objects and HAeBe stars, respectively.
This trend is also seen in the case of \methanol\ masers.
The detection rate of \methanol\ masers toward Class~0 objects is $36~\%$, 
which is much higher than those of Class~I ($21~\%$) and HAeBe ($1~\%$) sources.

It should be noted that the detection rates of \water\ and 44~\ghz\ \methanol\ masers 
for HAeBe stars and candidates in our survey might be lower limits for bona fide intermediate-mass PMS stars.
Although the classification criteria introduced in Section 1 were intended 
to identify IMYSOs,
some of the HAeBe candidates in our sample may not be in the PMS phase.
The classification criteria allow contamination with objects
such as post-asymptotic giant branch stars and protoplanetary nebulae,
which could exhibit similar observational features
(\citealt{The94, Sartori10, Vieira11}). 
For example, \citet{Sartori10} tested the young nature of 93 HAeBe candidates 
selected from the sample of \citet{Vieira03},
and concluded that
at least 71 (76\%) sources are PMS stars but 7 sources are evolved stars.
Our sample contains 22 (15\%) sources with ambiguity in their classification.
They are indicated with question marks in Column 4 of Table~\ref{table:source}. 
None of them show \water\ or \methanol\ maser emission.
Even after accounting for this contamination, however,
the detection rates of both masers toward intermediate PMS stars would 
still be much lower than those of Class~0 and Class~I objects.

In this survey, we report the first detection of \water\ maser emission in
one source (HH~165) and of 44~\ghz\ \methanol\ maser emission in six sources
(CB~3, IRAS~00338+6312, OMC3~MMS9, IRAS~05338$-$0624, V1318~CygS, and IRAS~23011+6126).
We focus on two HAeBe stars, HH~165 and V1318~CygS, 
because {\it no} HAeBe star has previously been found to show 44~\ghz\ \methanol\ maser emission and {\it only eight} HAeBe stars have been reported to show 22~\ghz\ \water\ maser emission:
LkH$\alpha$~198, HD 250550, V373 Cep \citep{Schwartz75}, IRAS 06571$-$0441 \citep{Han98},
V1318 CygS \citep{Palla95}, PV Cep \citep{Torrelles86}, V1057 Cyg \citep{Rodriguez87}, and V645 Cyg \citep{Lada81}.
Table \ref{table:HAeBe} lists the eight with the detected line parameters.
We observed all of them but did not detect any appreciable ($>$1~Jy)
\water\ and \methanol\ maser emission from four of them: LkH$\alpha$~198, HD~250550, PV~Cep, and V1057~Cyg.

\subsection{Notes on Some Individual Sources}

As mentioned in the previous section, some sources are located 
within a single beam or first sidelobe so that 
it is not straightforward to interpret the observational results.
In order to determine genuine maser-emitting source(s) in such cases,
we observed closely located objects successively and examined 
the detected maser lines of the individual sources.
In the process, we assumed that the main beam of the telescope is Gaussian
and used the FWHMs listed in Table \ref{table:obs}, 130$''$ at 22~\ghz\ and 65$''$ at 44~\ghz.

\subsubsection{IRAS~05338$-$0647 and HH147~MMS}

\begin{figure*}
 \centering
 \includegraphics{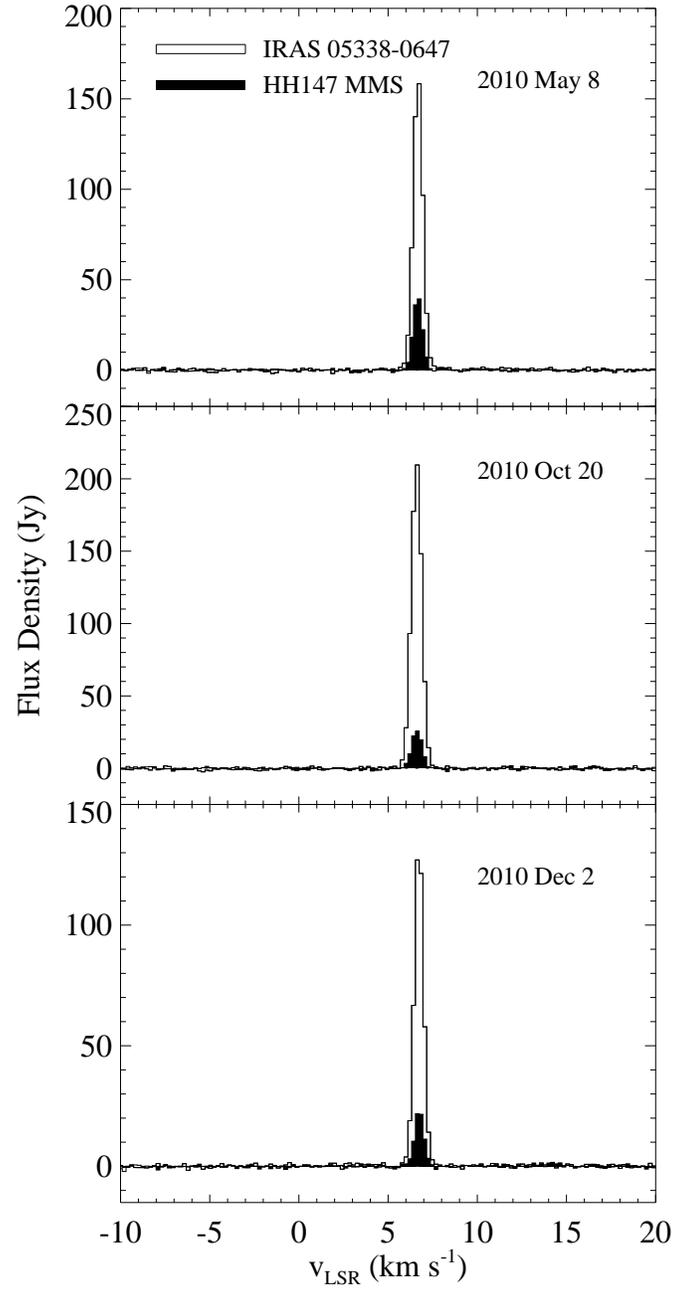}
 \caption{Spectra of the \water\ maser emission detected toward IRAS~05338-0647 and HH147~MMS.\label{fig:HH147}}
\end{figure*}

\begin{figure*}
 \centering
 \includegraphics{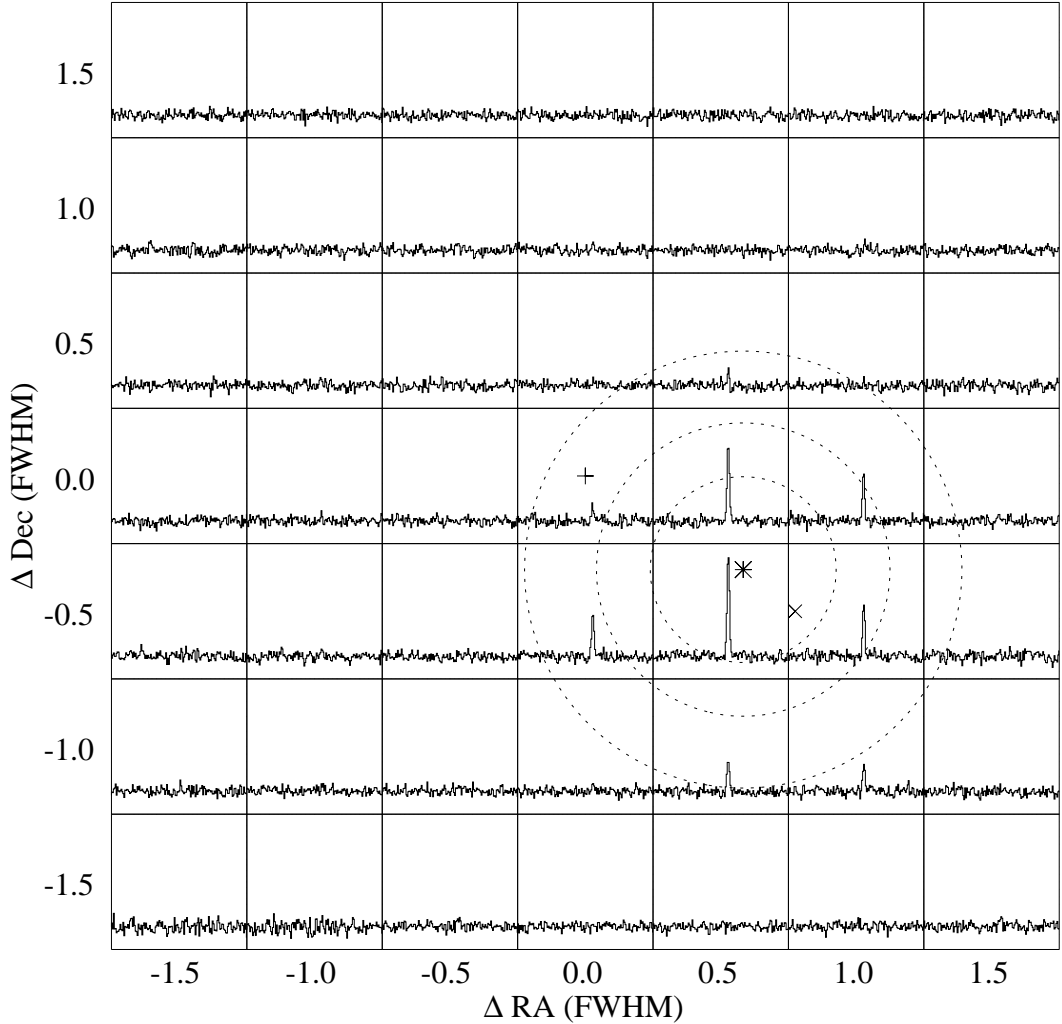}
 \caption{Grid mapping results of \water\ maser emission from HH147~MMS at the center (0.0, 0.0).
The intensity pattern is well fitted by a two-dimensional Gaussian model
with a similar FWHM to that presented in Table 2. 
 The result is shown as dotted contours.
 Contour levels are 25\%, 50\%, and 75\% of the Gaussian peak.
 The position of HH147~MMS is indicated with a plus symbol, 
 while that of IRAS~05338$-$0647 is marked by a cross symbol.
 The asterisk symbol represents the position of the Gaussian peak.
In each panel the horizontal and vertical axes are $v_{\rm LSR}$ and flux density,
 respectively. The $v_{\rm LSR}$ range is ($-10$ to +20)~\kms,
 while the flux density range is ($-$30 to +220)~Jy.
\label{fig:HH147_grid}}
\end{figure*}

IRAS~05338$-$0647 coincides with the ridge that contains HH~1$-$2 MMS~2$-$3 
while HH147~MMS ($\equiv$ IRAS~05339$-$0646) corresponds to a YSO, 
which drives the HH147 bipolar outflow (\citealt{Strom85, Eisloffel94, Chini01}).
These two objects are $121''$ apart from each other.
Figure \ref{fig:HH147} displays the single \water\ maser lines detected toward these sources over three epochs.
The two maser lines have practically the same velocity and line profile in each epoch.
This strongly suggests that they are not distinct maser sources but rather are a single maser.  
However, it should be noted that the observed peak intensity ratio of
HH147~MMS to IRAS~05338$-$0647 changes considerably over the observations,
25\% in 2010 May, 12\% in October, and 17\% in December.
If the maser is variable but stays at the same position,
the peak intensities of the two objects can vary 
but the ratio should be constant.
These large intensity ratio variations can be explained only by $\gtrsim10''$ pointing offsets
even though off-center pointing is considered.
It is also very {\it unlikely} that
the large ratio variations were caused by 
movement of a single maser,
because the typical proper motion of a maser observed in
other star-forming regions is only $\sim0.''01~\yr$ (e.g., \citealt{Hirota07}).
Therefore, it is more plausible that the maser emission in the latter epoch 
was emitted from new maser source formed after the disappearance of 
the maser source in the former epoch.
Taking the intensity ratios into account, the masers seem to 
be located between the two objects, much closer to IRAS~05338$-$0647.
This is consistent with the result of the grid mapping presented in Figure \ref{fig:HH147_grid},
which was made in the first epoch to investigate the existence of 
a nearby strong maser source (see Section 3).

\subsubsection{V1685~Cyg, V1686~Cyg, and V1318~CygS}

BD+40\degree\ is an active star-forming region, 
which contains at least 33 optical and NIR sources in a $\sim 1 ' \times 2'$ field, including 3 HAeBe stars:
V1685~Cyg ($\equiv$ BD+40\degree4124), V1686~Cyg ($\equiv$ LkH$\alpha$~224), and V1318~CygS ($\equiv$ LkH$\alpha$~225S) \citep{Hillenbrand95}.
Using the Medicina 32~m telescope,
\citet{Palla93} first detected \water\ maser emission towards 
BD+40\degree\ and claimed the maser emission stems from V1685~Cyg.
However, \citet{Palla95} observed this region in the \water\ maser
line at $\sim0."1$ spatial resolution using the Very Large Array (VLA),
and found that the maser is coincident with V1318~CygS
rather than V1685~Cyg.
\citet{Marvel05} also detected \water\ maser emission toward V1318~CygS, 
in the $v_{\rm LSR}$ range between $-$80~\kms\ and +20~\kms, 
using the VLA and the Very Long Baseline Array.

\begin{figure*}
 \centering
 \includegraphics[width=\textwidth]{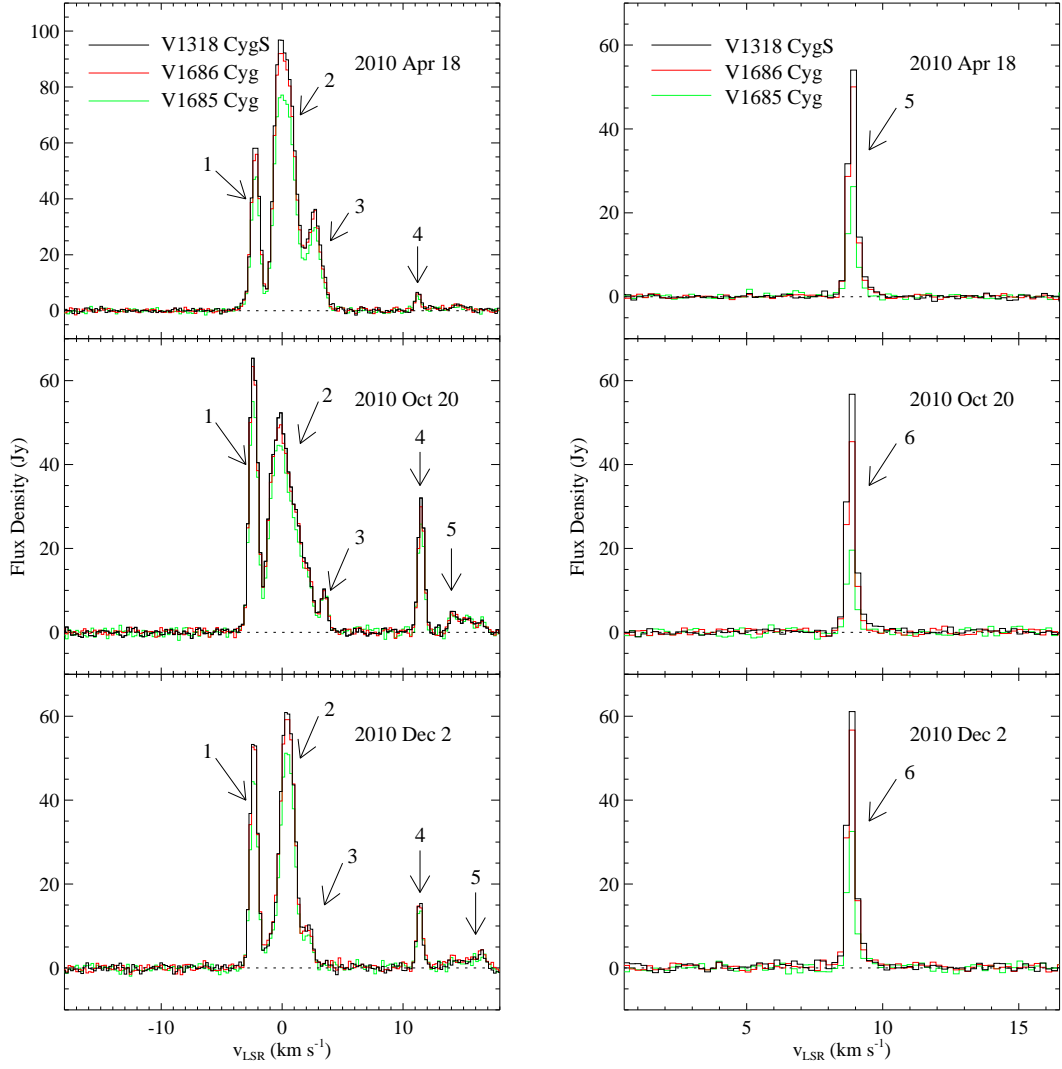}
 \caption{Left: spectra of the \water\ maser emission detected toward three HAeBe stars located in the BD+40\degree\ region, 
 V1318~CygS (black lines), V1686~Cyg (red lines), and V1685~Cyg (green lines).
 Right: spectra of the \methanol\ maser emission.\label{fig:BD40}}
\end{figure*}

We observed all three HAeBe stars on the same days and detected both \water\ and \methanol\ 
maser lines toward all of them (Figure \ref{fig:BD40}).
Assuming that the maser emission emanates from V1318~CygS,
we estimated the expected line intensities
at V1685~Cyg and V1686~Cyg, which are $36''$ and $14''$
from V1318~CygS on the sky, respectively.
If the pointing accuracy is $5''$, $\sim~81~\%\pm5~\%$ at 22~\ghz\ and
$\sim43~\%\pm10~\%$ at 44~\ghz\ of the original signal
would be detected toward V1685~Cyg.
In the case of V1686~Cyg, 
the values are $\sim97~\%\pm2~\%$ lower at 22~\ghz\ and
$\sim88~\%\pm8~\%$ at 44~ \ghz.
Table \ref{table:BD40} displays the observed and expected intensities.
The observed intensities of the individual \water\ and \methanol\
maser lines are in very good agreement with the expected intensities for all three observations.
This strongly suggests that all the \methanol\ maser lines detected toward
these three HAeBe stars 
originate from V1318~CygS.
Interferometric observations will be required to clarify this issue.

\subsubsection{NGC~7129~FIRS2, V373~Cep, and V361~Cep}

NGC~7129 is a reflection nebula illuminated by several stars including one
Class~0 object, NGC~7129~FIRS2 \citep{Eiora98}, 
and two HAeBe stars, V361~Cep and V373~Cep ($\equiv$ LkH$\alpha$~234).
\citet{Cesarsky78} detected \water\ maser emission toward
NGC~7129~FIRS2 and V373~Cep.
However, we detected \water\ maser emission toward V361~Cep
as well as the other two sources.
The separation angles of V361~Cep from NGC 7129~FIRS2
and V373~Cep are $201''$ and $102''$, respectively.
Since the KVN telescope has the first sidelobe
with an attenuation of 14~dB at $\sim$1.5 times FWHM away
from the main beam center,
as noted earlier, $\sim3.9~\%\pm0.2~\%$ of the \water\ maser emission from NGC~7129~FIRS2
and $\sim18~\%\pm3~\%$ of the \water\ maser emission from V373~Cep
would be detected toward V361~Cep. 
Table \ref{table:NGC7129} presents both the observed maser line intensities of the three objects and 
the theoretically expected values.
They match well with each other.
Therefore, it seems that the \water\ maser emission arises only from
NGC~7129~FIRS2 and V373~Cep.

\section{DISCUSSION}

\subsection{Detection Rates}

As mentioned in Section 4.1, the overall detection rates of \water\ 
and \methanol\ masers in this study are $9~\%$ and $6~\%$, 
respectively.
If only Class~0 and Class~I sources are included, 
the rates are 33~\% for \water\ and 27~\% for \methanol.
For comparison, \citet{Wilking94} made a multi-epoch \water\ maser 
line survey of
42 nearby ($d$~$<$~450~pc) Class~I sources with $L_{\rm FIR}$~$<$~120~$L_\odot$,
and detected maser emission in 36~\% of them.
\citet{Furuya01} also observed \water\ maser emission towards
30 Class~0 and 33 Class~I objects with $L_{\rm FIR}$~$<$~100~$L_\odot$. 
The detection rate was 21\%.
On the other hand,
\citet{Sridharan02} detected \water\ masers in 42\% of 84 
high-mass protostellar candidates.
\citet{Szymczak05} also detected 52\% in a sample of 79 Class~II \methanol\ maser sources at 6.7~GHz, which may trace an earlier stage of high-mass
star formation than UCHII regions. 
Both surveys were made with the Effelsberg 100~m telescope.
In the case of UCHII regions, whose ionizing stars have already reached the main sequence, the detection rate appears to be much higher.
\citet{Churchwell90} detected \water\ maser
emission in 67~\% of 84 UCHII regions with a similar flux limit
(mean rms $\sim$~0.2~Jy at a velocity resolution of $\sim$~0.2~\kms)
to those of \citet{Sridharan02} and \citet{Szymczak05},
using the same telescope.
This is contrary to the results of the low- and intermediate-mass regime
where the detection rates rapidly decrease after the central stars reach
the PMS stage. 
We detected \water\ maser emission in only 3~\% of 147 HAeBe stars 
(intermediate-mass PMS stars) and
\citet{Furuya01} detected none in a sample of 9 Class~II sources
(low-mass PMS stars).

A similar trend is seen in 44~\ghz\ Class~I \methanol\ masers.
\citet{Fontani10} recently made a survey of this maser line toward
88 high-mass protostellar candidates and obtained a detection rate of 31\%
at a similar detection limit to this study.
The sources in their sample are divided into two groups: $low$ and $high$.
The sources in the $high$ group may be at later evolutionary stages than
those in the $low$ group. 
The $high$ group has a detection rate about three times higher 
than the low group: 48\% versus 17\%. 
\citet{Haschick90} surveyed
50 star-forming regions, the vast majority of which are
ultracompact and compact HII regions.
They detected maser emission in 50\% of them
with one order of magnitude higher flux limit than this study.
In the high-mass regime, therefore, the detection rate of 44~\ghz\ 
Class~I \methanol\ maser emission also increases significantly 
as the central (proto)stars evolve (see \citealt{Voronkov10a}
for 9.9~\ghz\ Class~I \methanol\ masers). 
This is in contrast with the HAeBe stars, for which we find a 
detection rate of 1~\% for \methanol\ masers.

Consequently, taking into account that UCHII regions are still deeply 
embedded in dense molecular cores, whereas low- and intermediate-mass 
PMS stars have already emerged as visible objects,
the occurrence of both masers seems to be closely related to 
the circumstellar environments
as well as the evolutionary stage of the central objects (see Section 4.1).

\subsection{Relative velocities of masers with respect to the ambient gas}

While the details are still poorly understood, 
both \water\ and Class~I \methanol\ masers are thought to be collisionally pumped 
(\citealt{Elitzur89, Cragg92}).
The presence of shocks is a necessary condition to enhance the abundances of \water\ and \methanol\ molecules, 
favoring the formation of collisionally pumped masers.
One plausible mechanism for inducing interstellar shocks is jets and outflows from YSOs.
In fact, \water\ and Class~I \methanol\ masers were found to be
associated with jets and/or outflows in many YSOs 
(\citealt{Felli92, Kurtz04, Cyganowski09}),
although class~I \methanol\ maser emission can be related to 
expanding HII~regions \citep{Voronkov10a}.

Figure \ref{fig:velocity} displays the relative velocities of both masers 
with respect to the natal dense molecular cores against the bolometric luminosities of the central (proto)stars, 
$L_{\rm bol}$. 
Here the peak velocities ($v_{\rm peak}$) of Tables \ref{table:P22} and \ref{table:P44} are used.
The molecular gas velocities are collected from the literature and are presented in 
Table~\ref{table:parameter} with the references.
In Figure \ref{fig:velocity}(a), the velocity difference between
\water\ masers and the associated molecular gas 
has an average of 9.3~\kms\ and a median of 6.5~\kms.
While these values are much smaller than the typical shock velocity 
($\sim$~100~\kms) suggested by theoretical models \citep{Elitzur89},
they are similar to the observed velocities of molecular outflows,
which range from a few \kms\ to about 20~\kms\ (e.g., \citealt{Kim06}), as well as 
the velocity differences obtained by the previous \water\ maser surveys, 
$\sim~5-10~\kms$ \citep{Churchwell90,Palla91,Kurtz05,Urquhart09}.
In addition, 
large relative velocities ($>10~\kms$) are mostly confined to objects in the earliest, 
Class~0, evolutionary stage.
This is probably because outflows are strongest during the Class~0 stage 
and are weakened as the central (proto)stars evolve (\citealt{Bontemps96, Arce06}).
On the other hand,
\methanol\ masers usually show much smaller relative velocities than \water\ masers.
In Figure \ref{fig:velocity}(b), their relative velocities are clustered around 0~\kms\ and do not deviate more than 10~\kms\ 
(see also \citealt{Bachiller90,Fontani10}).
Furthermore, the sources with relative velocities $>1~\kms$ are all Class~0 objects. 
This might also be accounted for with the strength of outflows, as for \water\ masers.
However, it is worth noting that there is at least one example of a Class~I \methanol\ maser
that shows a significant velocity offset ($\sim30~\kms$) from the associated 
molecular gas \citep{Voronkov10b}.

Previous high-resolution studies have shown that \water\ masers are 
usually located closer to the central (proto)stars 
than Class~I \methanol\ masers (e.g., \citealt{Kurtz04}).
It is thus widely believed that \water\ maser emission originates from the
inner parts of outflows while Class~I \methanol\ maser emission comes from
the interacting interface of outflows with the ambient gas.
This view is supported by theoretical studies showing 
that \water\ masers form in warm ($\sim500$~K), very dense 
($\sim10^{9}~{\rm{cm}}^{-3}$) gas behind high-velocity shocks
while Class~I \methanol\ masers can arise in much less dense
($\sim10^{5}~{\rm{cm}}^{-3}$) postshock gas 
at considerably lower ($\sim100$~K) temperatures
(\citealt{Elitzur89, Cragg92}).
Moreover, the \methanol\ molecule can survive sputtering or desorption of grain mantles
only at low ($<10~\kms$) shock velocities \citep{Garay02}.
Our study cannot distinguish different spatial distributions of 
\water\ and \methanol\ masers owing to low angular resolution.
Nevertheless, our spectroscopic results indicate that the emitting regions of the two masers are different. 
Our results support the previous suggestions that \water\ maser emission originates from the
base of an outflow whereas Class~I \methanol\ maser emission arises from
the interaction region of the outflow with the ambient gas.
Figure \ref{fig:velocity} also shows that no correlation exists between the relative velocities 
of either maser and the bolometric luminosities of the central objects.

\subsection{Variability of masers}

It is well known that
the intensity and shape of \water\ maser line emission can vary considerably
on timescales of weeks to years
(\citealt{Tofani95, Claussen96, Furuya03, Brand03, Breen10}).
Although our observations are confined to three epochs within a year,
Figure 1 shows notable changes in \water\ maser line intensity and velocity.
In particular, the integrated line intensity of Serpens~FIRS1 ($v_{\rm LSR}\simeq+23~\kms$) decreased by a factor of 70 in seven months,
and that of NGC~7129~FIRS2 ($v_{\rm LSR}\simeq-8~\kms$) increased by a factor of 30 in five months.
Moreover, the appearance of new maser lines (CB~3, Serpens~FIRS1, IRAS~20050+2720~MMS1, and S106~FIR) and 
disappearance of others (IRAS~00338+6312, IRAS~05375$-$0731, IRAS~06571$-$0441, IRAS~20050+2720~MMS1, and IRAS~23037+6213) 
is further evidence of variability.

Despite the short lifetime of the \water\ maser lines and the blending of several velocity components, 
we could find line-of-sight velocity drifts of \water\ maser lines 
in half of the 16 maser-detected sources:
CB~3, IRAS~00338+6312, HH~165, IRAS~20050+2720~MMS1, V1318~CygS, V645~Cyg, IRAS~21391+5802, and NGC~7129~FIRS2.
The velocity gradients of these sources except one are $0.5-2.2$ \kms\ \yr,
which are similar to other published results, e.g.,
1.5~\kms~\yr\ \citep{Hunter94}, 1.2~\kms~\yr\ \citep{Tofani95}, and $<$~1.8~\kms~\yr\ \citep{Brand03}.
The exception is IRAS~20050+2720~MMS1, which shows an extremely large velocity gradient of $6.0~\kms~\yr$.
Because of the coarse sampling and the low spatial resolution of our observations, however, we cannot exclude the possibility that the velocity variation was caused by different maser features, i.e., the disappearance of one maser feature and the appearance of another at a similar position.

On the other hand, the variability of the 44~\ghz\ \methanol\ maser has not been well established.
By comparing their observational results with the published data, 
\citet{Kurtz04} could not find \methanol\ maser variability except for two objects in their sample.
\citet{Kalenskii10} presented observational results of three low-mass YSOs at three epochs with two-year intervals.
Their spectra do not reveal any significant variability.
Our results exhibit only small variation in the \methanol\ maser line
intensity, shape, and velocity over a one-year interval (Figure \ref{fig:methanol}).
The integrated intensities are changed by $\lesssim50~\%$ except for 
IRAS~05338-0624 and IRAS~23011+6126, which have poor signal-to-noise ratios.

\citet{Tofani95} found from the \water\ maser line observations of 22 YSOs with the VLA and Medicina 32~m radio telescopes that 
variability of the maser emission tends to be
more pronounced for maser spots closer to the central star and for those
with larger relative velocity with respect to the molecular gas. 
Thus, these different variability behaviors of \water\ and
44~\ghz\ \methanol\ masers
may also be connected with different emitting environments (see Section 5.2).

\subsection{The relationship of maser luminosity with bolometric luminosity}

The isotropic maser luminosity can be calculated from the observed line
integral using the following equations: 

\begin{eqnarray}
L_{\rm{H_2O}} & = &4 \pi d^{2} {\frac{\nu}{c}} \int{F_{\nu}}~dv 
\nonumber \\
& = & 5.79 \times 10^{-9} L_{\odot} ~\bigg(\frac{d}{500\rm{pc}}\bigg)^2 \bigg(\frac{\int{F_{\nu}}~dv}{1~\rm{Jy~km~s}^{-1}}\bigg) \bigg(\frac{\eta_{a}}{0.72}\bigg)^{-1}
\end{eqnarray}

and

\begin{equation}
L_{\rm{CH_3OH}} = 
1.15 \times 10^{-8} L_{\odot} ~\bigg(\frac{d}{500\rm{pc}}\bigg)^2 \bigg(\frac{\int{F_{\nu}}~dv}{1~\rm{Jy~km~s}^{-1}}\bigg) \bigg(\frac{\eta_{a}}{0.69}\bigg)^{-1}.
\end{equation}

\noindent
Here $d$ is the distance to the source, 
$\eta_a$ is the antenna efficiency at the observed frequency $\nu$, 
and $A_{p}$ is the geometric area of the antenna aperture plane.
Table~\ref{table:parameter} presents the derived isotropic maser luminosities
using the distances of the central objects presented in Table~\ref{table:source}. 
For sources with multiple maser lines, the maser luminosity 
is calculated as the sum of all the individual lines. 

Figure~\ref{fig:lum}(a) plots $L_{\rm{H_2O}}$ against $L_{\rm bol}$.
There is a good correlation between the two.
A linear least-squares fit to our observed data points results in 
$L_{\rm{H_{2}O}} = 1.71 \times 10^{-9} (L_{\rm{bol}})^{0.97}$
with a correlation coefficient ($\rho$) of 0.72.
Figure~\ref{fig:lum}(b) exhibits a plot of $L_{\rm{CH_3OH}}$ versus 
$L_{\rm bol}$.
The fitted relation is 
$L_{\rm{CH_{3}OH}} = 1.71 \times 10^{-10} (L_{\rm{bol}})^{1.22}$
with $\rho$=0.71.
The slope of 1.22 is steeper than that for the \water\ masers.
It is expected from the two relations
that there may be a correlation between $L_{\rm{H_{2}O}}$ and 
$L_{\rm{CH_{3}OH}}$.
Figure \ref{fig:lum}(c) shows the relation.
A linear fit to the data points yields 
$L_{\rm{CH_{3}OH}} = 1.13 \times 10^{-4} (L_{\rm{H_{2}O}})^{0.42}$ ($\rho$=0.45).

For comparison, we plot 20 \water\ maser data points of low-mass ($L_{\rm{bol}} < 40~L_{\odot}$) YSOs 
and 38 \water\ and 30 \methanol\ maser data points of UCHII regions 
together with our data in Figure \ref{fig:lum_all} (\citealt{Furuya03}; Kim et al. 2011, in preparation).
The data of UCHII regions were obtained with the same telescope as in this survey.
One can clearly see that the luminosities of both masers
are well correlated with the bolometric luminosity. 
The fitted relations are $L_{\rm{H_{2}O}} = 4.10 \times 10^{-9} (L_{\rm{bol}})^{0.84}$ ($\rho$=0.88)
and $L_{\rm{CH_{3}OH}} = 2.89 \times 10^{-9} (L_{\rm{bol}})^{0.73}$ ($\rho$=0.80), 
which are shown by the solid lines in Figures \ref{fig:lum_all}(a) and (b).
Figure \ref{fig:lum_all}(c) shows $L_{\rm{CH_{3}OH}}$ versus $L_{\rm{H_{2}O}}$ 
with the fitted relation,
$L_{\rm{CH_{3}OH}} = 3.44 \times 10^{-3} (L_{\rm{H_{2}O}})^{0.61}$ ($\rho$=0.73).

Several previous studies of \water\ masers in Galactic star-forming
regions have produced similar relations
of $L_{\rm H_2O}$ with $L_{\rm bol}$ 
(\citealt{Wouterloot86, Felli92, Brand03, Furuya03}).
Our slope agrees well with the previous values, 0.8$-$1.0.
For a given evolutionary stage, therefore, YSOs with higher bolometric 
luminosities are expected to have higher \water\ maser luminosities.
However, we are unaware of any study examining the $L_{\rm bol}$-$L_{\rm{CH_{3}OH}}$ and $L_{\rm H_2O}$-$L_{\rm{CH_{3}OH}}$ relations.

\section{SUMMARY}

We have carried out a multi-epoch, simultaneous 22~\ghz\ \water\ 
and 44~\ghz\ Class I \methanol\ maser line survey toward 
180 IMYSOs.
The main results are summarized as follows. 

1. We detected \water\ masers toward 16 objects and \methanol\ masers toward 10 objects.
One new \water\ maser source (HH~165) and six new \methanol\ maser sources 
(CB~3, IRAS~00338+6312, OMC3~MMS9, IRAS~05338$-$0624, V1318~CygS, and IRAS~23011+6126) 
were found in our survey.

2. The overall detection rates of \water\ masers and \methanol\ masers are $9~\%$ and $6~\%$, respectively.
The rates rapidly decrease as the central (proto)stars evolve.
The detection rates of \water\ masers are $50~\%$, $21~\%$ and, $3~\%$ for Class~0, Class~I, and HAeBe objects, respectively.
Those of \methanol\ masers for Class~0, Class~I, and HAeBe objects are $36~\%$, $21~\%$, and $1~\%$.
In contrast, the detection rates of both masers in high-mass star-forming regions significantly increase 
as the central objects evolve from the protostellar to the main-sequence stage.
These results indicate that the occurrence of the two masers are closely related both to
the evolutionary stage of the central (proto)stars and to the circumstellar environments. 

3. The relative velocities of \water\ masers with respect to
the ambient molecular gas are 9.3~\kms\ on average, with a median difference of 6.5~\kms, 
whereas those of \methanol\ masers are concentrated around 0 \kms.
No \methanol\ maser velocity deviates more than 10~\kms\ from the systemic velocity.
Large relative velocities are mainly shown in the Class~0 objects: 
$\left| v_{\rm{H_{2}O}} - v_{\rm{sys}} \right| > 10~\kms$ and 
$\left| v_{\rm{CH_{3}OH}} - v_{\rm{sys}} \right| > 1~\kms$.
This is consistent with previous suggestions that \water\ masers originate from the 
inner parts of outflows while Class~I \methanol\ masers arise from the interacting interface of 
outflows with the ambient dense gas.

4. The intensities and shapes of the observed \water\ maser lines
were quite variable. Half of the maser-detected sources show
velocity drifts. The integrated line intensities varied by up to 
two orders of magnitude.
In contrast, the observed \methanol\ lines do not reveal any significant
variability in intensity, shape, or velocity. 
The line integrals were maintained within $\sim50~\%$ over the observations.
These different variability behaviors of the two masers may be connected with
different emitting environments.

5. The isotropic luminosities of both masers are well correlated
with the bolometric luminosities of the central objects.
The linear fits result in 
$L_{\rm{H_{2}O}} = 1.71 \times 10^{-9} (L_{\rm{bol}})^{0.97}$
and 
$L_{\rm{CH_{3}OH}} = 1.71 \times 10^{-10} (L_{\rm{bol}})^{1.22}$
when only IMYSO data in this survey were considered,
while those yield
$L_{\rm{H_{2}O}} = 4.10 \times 10^{-9} (L_{\rm{bol}})^{0.84}$
and
$L_{\rm{CH_{3}OH}} = 2.89 \times 10^{-9} (L_{\rm{bol}})^{0.73}$,
after the data points of low- and high-mass regimes are added.

\acknowledgments
We are very grateful to Stan Kurtz for carefully reading the manuscript
and for many helpful comments. We also thank the anonymous referee
for constructive comments.

\clearpage

\clearpage

\clearpage

\clearpage

\clearpage

\clearpage
\begin{figure*}
 \centering
 \includegraphics{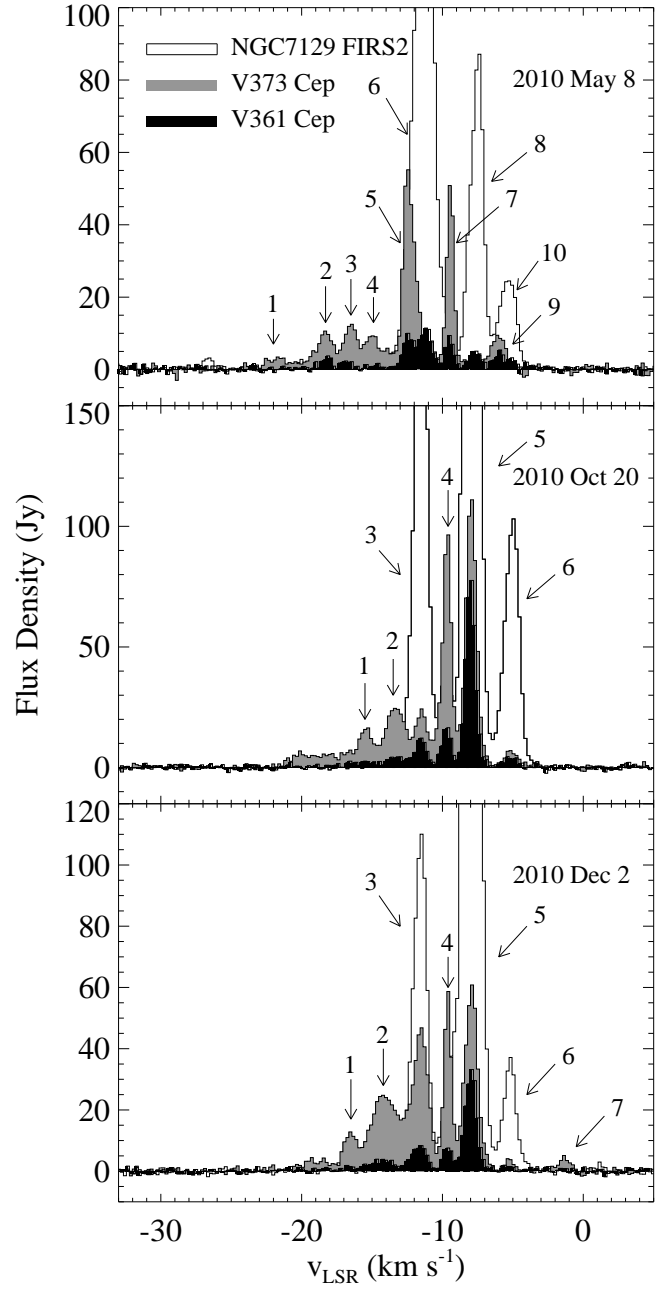}
 \caption{Spectra of the \water\ maser emission detected toward NGC7129~FIRS2, V373~Cep, and V361~Cep. 
 \label{fig:NGC7129}}
\end{figure*}

\clearpage
\begin{figure*}
 \includegraphics[width=\textwidth]{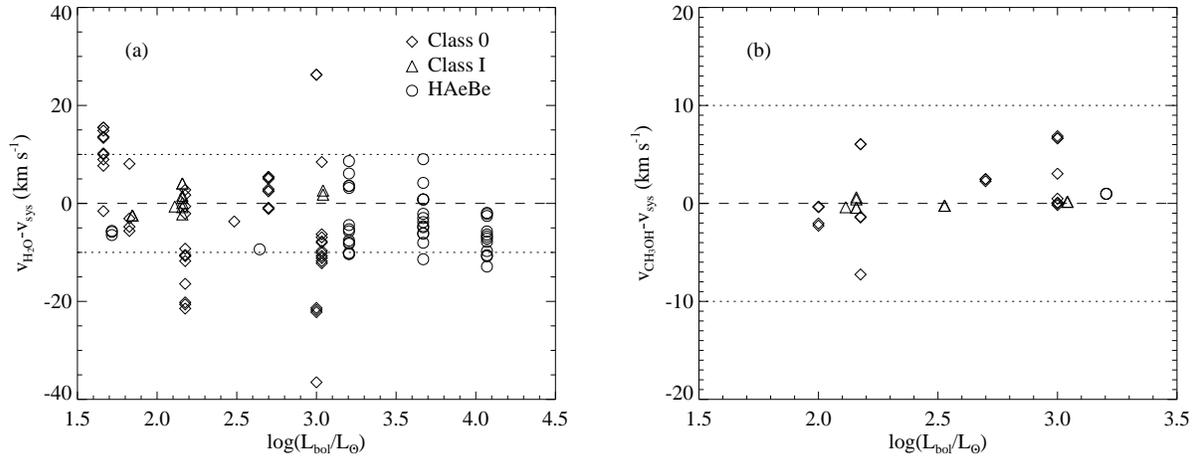}
 \caption{Relative velocity with respect to the associated molecular gas vs. the bolometric luminosity of the central (proto)star for (a) \water\ maser lines and (b) 44~\ghz\ Class~I \methanol\ maser lines. 
 Diamonds, triangles, and circles represent data for Class~0, Class~I, and HAeBe sources, respectively.
 The dashed lines represent the zero velocity difference and the dotted lines indicate $10~\kms$ of velocity difference.
This figure does not include data of IRAS~20050+2720~MMS1, which shows an
exceptionally large velocity difference, $-90~\kms$ in the first observation and $-93~\kms$ in the second observation.
 \label{fig:velocity}} 
\end{figure*}

\clearpage
\begin{figure*}
 \includegraphics[width=\textwidth]{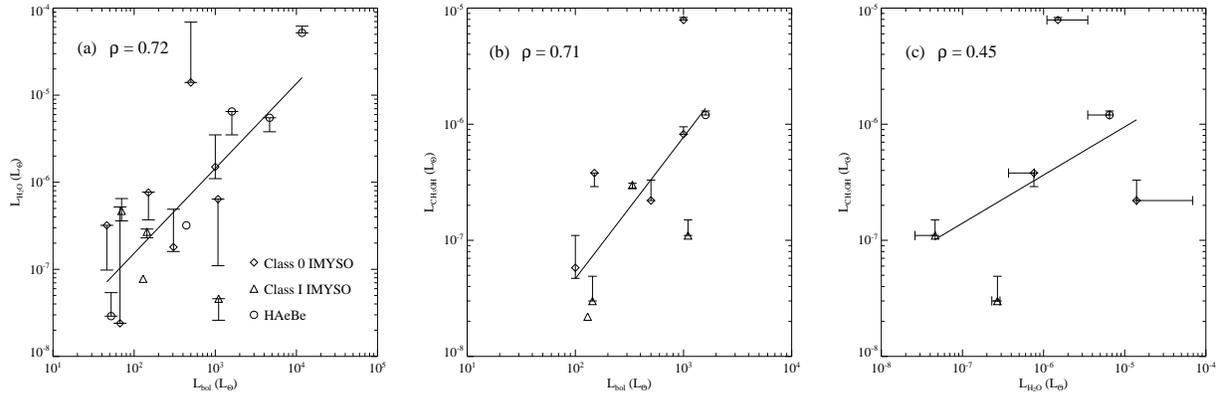}
 \caption{(a) \water\ maser luminosity vs. bolometric luminosity,
 (b) \methanol\ maser luminosity vs. bolometric luminosity, and 
 (c) \methanol\ maser luminosity vs. \water\ maser luminosity.
 Diamonds, triangles, and circles in each panel indicate the data points of
 Class 0, Class I, and HAeBe sources in the first epoch, respectively.
 The error bars represent the variability of maser luminosity through the entire observations.
In each panel, the solid line is the fitted relation to the data points
and the correlation coefficient is shown at the upper left corner.\label{fig:lum}}
\end{figure*}

\clearpage
\begin{figure*}
 \includegraphics[width=\textwidth]{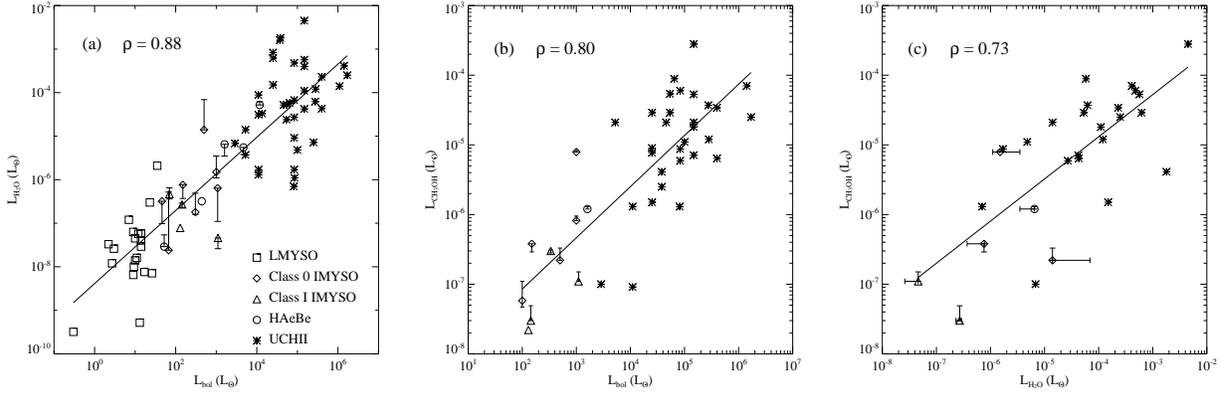}
 \caption{Same as Figure \ref{fig:lum}, but data in low- and high-mass regimes are added.
 Squares in panel (a) are the data points of low-mass YSOs from \citet{Furuya03}, 
 while asterisks in all three panels represent the data points of UCHII regions from Kim et al. (2011, in preparation).
In each panel the solid line is the fitted relation to all data points.
 \label{fig:lum_all}}
\end{figure*}

\clearpage
\begin{deluxetable}{lcccccccclccccc}
\tablecolumns{20}
\rotate
\tabletypesize{\tiny}
\tablecaption{
Source Summary\label{table:source}}
\tablewidth{0pt}
\tablehead{
\colhead{} & 
\colhead{R.A.} &
\colhead{Decl.} &
\colhead{} &
\colhead{} &
\colhead{Distance} &
\colhead{} &
\colhead{$L_{\rm{bol}}$} &
\colhead{} &
\colhead{Observing} &
\multicolumn{2}{c}{Detection} &
\colhead{$\sigma_{\water}$} &
\colhead{$\sigma_{\methanol}$} \\
\cline{11-12}
\colhead{Source} & 
\colhead{(J2000)} & 
\colhead{(J2000)} & 
\colhead{Source Type\tablenotemark{a}} &
\colhead{Ref.} &
\colhead{(pc)} &
\colhead{Ref.} &
\colhead{($L_{\odot}$)} &
\colhead{Ref.} &
\colhead{Date} &
\colhead{\water} &
\colhead{\methanol} &
\colhead{(Jy)} & 
\colhead{(Jy)} \\
\colhead{(1)} & 
\colhead{(2)} &
\colhead{(3)} & 
\colhead{(4)} & 
\colhead{(5)} &
\colhead{(6)} &
\colhead{(7)} &
\colhead{(8)} &
\colhead{(9)} &
\colhead{(10)} &
\colhead{(11)} &
\colhead{(12)} &
\colhead{(13)} &
\colhead{(14)} 
}
\startdata
MacC H12                      & 00 07 02.6 & +65 38 38 & HAeBe(?) & 1, 2       &  845     &  3       &  47.1    &  3 			 & 2010 Feb 24 & N & N &         0.46  & 0.52 \\
HBC 324                       & 00 07 30.7 & +65 39 52 & HAeBe    & 1, 2       &  850     &  4       &  60      &  4 			 & 2010 Mar 18 & N & N &         0.36  & 0.47 \\ 
LkH$\alpha$ 198               & 00 11 26.0 & +58 49 29 & HAeBe    & 1, 2       &  600     &  4       &  115     &  4 		 	 & 2010 Feb 24 & N & N &         0.46  & 0.55 \\
V376 Cas                      & 00 11 26.1 & +58 50 04 & HAeBe(?) & 1, 2       &  630     &  5       &  437     &  5	 		   & 2010 Feb 24 & N & N  &      0.49  & 0.51 \\
CB 3                          & 00 28 42.7 & +56 42 07 & Class 0  & 6          &  2500    &  6       &  1000    &  6 			 & 2010 May 8 & Y & Y\tablenotemark{b} &     0.45  & 0.53 \\
VX Cas                        & 00 31 30.7 & +61 58 51 & HAeBe    & 1, 2       &  619     &  7       &  31      &  7   		 & 2010 Feb 24 & N & N &         0.43  & 0.46 \\
IRAS 00338+6312               & 00 36 47.5 & +63 29 02 & Class I  & 8          &  850     &  8       &  1100    &  8 			 & 2010 Jan 18 & Y & Y\tablenotemark{b} &    0.32  & 0.41 \\
V594 Cas                      & 00 43 18.3 & +61 54 40 & HAeBe(?) & 1, 2       &  650     &  5       &  1950    &  5 			 & 2010 Mar 18 & N & N &         0.36  & 0.48 \\
\enddata
\tablenotetext{a}{Sources with a question mark were classified as HAeBe candidates initially, 
but later analysis suggests they are not in the HAeBe stage.}
\tablenotetext{b}{Newly detected objects in our survey.}
\tablenotetext{c}{\water~maser emission lines toward the position of NGC1333 IRAS2 were detected, but turned out to be the side lobe effect by HH7-11C.}
\tablenotetext{d}{\water~maser emission lines toward the position of OMC3 MMS6, OMC3 MMS9, OMC3 MMS7, OMC2 FIR4, and T Ori were detected, but turned out to be the side lobe effect by Orion KL.}
\tablerefs{(1) \citealt{The94}; (2) \citealt{Hernandez04}; (3) \citealt{Mannings00}; (4) \citealt{Hernandez04}; (5) \citealt{Manoj06}; 
(6) \citealt{Alonso10}; (7) \citealt{Montesinos09}; (8) \citealt{Furuya03}; (9) \citealt{Froebrich05}; (10) \citealt{Chen09}; 
(11) \citealt{Vieira03}; (12) \citealt{Sartori10}; (13) \citealt{Palla93}; (14) \citealt{Vieira11}; (15) \citealt{Yang02}; 
(16) \citealt{Malfait98}; (17) \citealt{Saraceno96}; (18) \citealt{Beltran01}; (19) \citealt{Dent98}; (20) \citealt{Chini01}; 
(21) \citealt{Wu96}; (22) \citealt{Sandell99}; (23) \citealt{Connelley08}; (24) \citealt{Gregorio92}; (25) \citealt{Wolf03}; 
(26) \citealt{Torres95}; (27) \citealt{Han98}; (28) \citealt{Mora01}; (29) \citealt{Dent92}; (30) \citealt{Furuya05}; 
(31) \citealt{Palla95}; (32) \citealt{Wu04}; (33) \citealt{Connelley07}.}
\end{deluxetable}

\clearpage
\begin{deluxetable}{ccccccc}
\tablecolumns{7}
\tabletypesize{\scriptsize}
\tablecaption{Description of The Observations\label{table:obs}}
\tablewidth{0pt}
\tablehead{
\colhead{} & 
\colhead{Frequency} &
\colhead{FWHM} &
\colhead{} &
\colhead{Velocity Resolution} &
\colhead{Velocity Coverage} & 
\colhead{System Temperature} \\
\colhead{Maser} & 
\colhead{(\ghz)} &
\colhead{($''$)} &
\colhead{$\eta_{\rm{MB}}$} &
\colhead{(\kms)} &
\colhead{(\kms)} & 
\colhead{(K)}
}
\startdata
\water\ $6_{16}-5_{23}$ & 22.23508 & 130 & 0.72 & 0.105 & 432 & 60 - 150 \\
\methanol\ $7_{0}-6_{1} A^{+}$ & 44.06943 & 65 & 0.69 & 0.053 & 218 & 120 - 210
\enddata
\end{deluxetable}

\clearpage
\begin{deluxetable}{llccccccccc}
\tablecolumns{13}
\tabletypesize{\tiny}
\tablecaption{\water\ Maser Line Parameters of Detected Sources\label{table:P22}}
\tablewidth{0pt}
\tablehead{
\colhead{} & 
\colhead{} &&
\multicolumn{3}{c}{Gaussian Fits} && 
\multicolumn{4}{c}{Integrated Lines} \\
\cline{4-6} \cline{8-11} 
\colhead{} & 
\colhead{Observing} &&
\colhead{$\int{F_{\nu}~dv}$} &
\colhead{$v_{\rm LSR}$} &
\colhead{$\delta v$} &&
\colhead{$\int{F_{\nu}~dv}$} &
\colhead{$F_{\rm peak}$} &
\colhead{$v_{\rm peak}$} &
\colhead{Equivalent Width} \\
\colhead{Source} & 
\colhead{Date} && 
\colhead{(Jy \kms)} & 
\colhead{(\kms)} & 
\colhead{(\kms)} && 
\colhead{(Jy \kms)} & 
\colhead{(Jy)} & 
\colhead{(\kms)} & 
\colhead{(\kms)} \\
\colhead{(1)} & 
\colhead{(2)} && 
\colhead{(3)} & 
\colhead{(4)} & 
\colhead{(5)} && 
\colhead{(6)} & 
\colhead{(7)} & 
\colhead{(8)} & 
\colhead{(9)}
}
\startdata
CB 3                 & 2010 May 7  &&   10.54$\pm$0.33  & $-60.24\pm0.01$ & 0.76 $\pm$ 0.03 &&  11.21 &   12.65&$-$60.15  &   0.88 \\
                     & 2010 Oct 20 &&   13.20$\pm$0.55  & $-60.74\pm0.03$ & 1.43 $\pm$ 0.09 &&  12.09 &    9.10&$-$60.57  &   1.32 \\
                     &             &&    4.10$\pm$0.44  & $-12.65\pm0.04$ & 0.66 $\pm$ 0.08 &&   4.21 &    5.32&$-$12.53  &   0.78 \\          
                     & 2010 Dec 2  &&    1.88$\pm$0.33  & $-75.13\pm0.09$ & 1.17 $\pm$ 0.23 &&   1.77 &    1.77&$-$75.31  &   0.99 \\
                     &             &&    4.55$\pm$0.44  & $-61.16\pm0.09$ & 2.12 $\pm$ 0.20 &&   4.21 &    2.10&$-$60.99  &   2.04 \\          
                     &             &&    1.77$\pm$0.11  & $-12.62\pm0.02$ & 0.43 $\pm$ 0.05 &&   1.88 &    3.33&$-$12.53  &   0.58 \\          
IRAS 00338+6312      & 2010 May 7  &&    2.77$\pm$0.33  & $-16.19\pm0.07$ & 1.16 $\pm$ 0.17 &&   2.99 &    2.44&$-$15.91  &   1.23 \\
                     & 2010 Oct 20 &&    1.55$\pm$0.33  & $-15.39\pm0.09$ & 0.60 $\pm$ 0.21 &&   1.66 &    2.44&$-$15.07  &   0.68 \\
\enddata
\end{deluxetable}

\clearpage
\begin{deluxetable}{llccccccccc}
\tablecolumns{13}
\tabletypesize{\tiny}
\tablecaption{\methanol\, Maser Line Parameters of Detected Sources\label{table:P44}}
\tablewidth{0pt}
\tablehead{
\colhead{} & 
\colhead{} &&
\multicolumn{3}{c}{Gaussian Fits} && 
\multicolumn{4}{c}{Integrated Lines} \\
\cline{4-6} \cline{8-11} 
\colhead{} & 
\colhead{Observing} &&
\colhead{$\int{F_{\nu}~dv}$} &
\colhead{$v_{\rm LSR}$} &
\colhead{$\delta v$} &&
\colhead{$\int{F_{\nu}~dv}$} &
\colhead{$F_{\rm peak}$} &
\colhead{$v_{\rm peak}$} &
\colhead{Equivalent Width} \\
\colhead{Source} & 
\colhead{Date} && 
\colhead{(Jy \kms)} & 
\colhead{(\kms)} & 
\colhead{(\kms)} && 
\colhead{(Jy \kms)} & 
\colhead{(Jy)} & 
\colhead{(\kms)} & 
\colhead{(\kms)} \\
\colhead{(1)} & 
\colhead{(2)} && 
\colhead{(3)} & 
\colhead{(4)} & 
\colhead{(5)} && 
\colhead{(6)} & 
\colhead{(7)} & 
\colhead{(8)} & 
\colhead{(9)}
}
\startdata
CB 3            & 2010 May 7  && 23.20 $\pm$ 0.92  & $-38.38\pm0.07$ & 3.82 $\pm$ 0.20 && 20.64  &  7.19  & $-$38.96 &  2.87  \\	  
                &             &&  4.87 $\pm$ 0.46  & $-32.01\pm0.06$ & 1.28 $\pm$ 0.17 &&  6.14  &  3.71  & $-$32.15 &  1.62  \\	  
                & 2010 Oct 20 && 18.09 $\pm$ 1.04  & $-38.52\pm0.06$ & 2.17 $\pm$ 0.15 && 18.21  &  8.58  & $-$38.32 &  2.12  \\
                &             &&  6.61 $\pm$ 1.27  & $-35.23\pm0.21$ & 2.52 $\pm$ 0.72 &&  4.06  &  4.75  & $-$35.77 &  0.88  \\	            
                &             &&  6.26 $\pm$ 0.92  & $-32.08\pm0.10$ & 1.48 $\pm$ 0.31 &&  5.56  &  4.98  & $-$32.15 &  1.11  \\	            
                & 2010 Dec 2  && 22.04 $\pm$ 0.81  & $-38.36\pm0.06$ & 3.27 $\pm$ 0.16 && 20.41  &  8.58  & $-$38.74 &  2.39  \\
                &             &&  8.46 $\pm$ 0.81  & $-32.14\pm0.08$ & 2.05 $\pm$ 0.26 &&  7.65  &  4.52  & $-$31.94 &  1.72  \\	            
IRAS 00338+6312 & 2010 May 7  &&  3.48 $\pm$ 0.34  & $-17.58\pm0.03$ & 0.78 $\pm$ 0.10 &&  3.24  &  4.06  & $-$17.54 &  0.81  \\
\enddata
\end{deluxetable}

\clearpage
\begin{deluxetable}{cccc}
\tablecolumns{4}
\tablecaption{Detection Rates of \water\ and \methanol\ Masers\label{table:detection}}
\tablewidth{0pt}
\tablehead{
\colhead{Source} & 
\colhead{Number} & 
\colhead{\water} & 
\colhead{\methanol} \\
\colhead{Type} & 
\colhead{of Sources} & 
\colhead{(\%)} & 
\colhead{(\%)} 
}
\startdata
Class~0    &        14         &        50 (7/14)     &           36 (5/14)  \\
Class~I    &        19         &        21 (4/19)     &           21 (4/19)  \\
HAeBe &       147         &         3 (5/147)    &            1 (1/147)  \\
   \hline
Total  &       180         &         9 (16/180)     &          6    (10/180)  \\   
\enddata
\end{deluxetable}

\clearpage
\begin{deluxetable}{llllrrc}
\tablecolumns{7}
\tablecaption{Previously Known Herbig Ae/Be Stars with \water\ Masers.\label{table:HAeBe}}
\tablewidth{0pt}
\tablehead{
\colhead{} & 
\colhead{R.A.} &
\colhead{Decl.} &
\colhead{Observing} &
\colhead{$v_{\rm{LSR}}$} &
\colhead{$F_{\rm peak}$} &
\colhead{} \\
\colhead{Source} & 
\colhead{(J2000)} & 
\colhead{(J2000)} & 
\colhead{Date} &
\colhead{(\kms)} & 
\colhead{(Jy)} & 
\colhead{Ref.}
}
\startdata
LkH$\alpha$ 198& 00 06 06\tablenotemark{a} & +58 17 00\tablenotemark{a} & 1974 Sep 6 &  $15$ & 3  & 1 \\
HD 250550  & 05 56 12\tablenotemark{a} & +16 31 00\tablenotemark{a} & 1975 Apr 20 & $13$      & 1.5& 1 \\
IRAS 06571$-$0441 & 06 57 06.4 & $-$04 41 48 & 1993 Nov 23 & $-23.5$     & 47.9 & 2 \\
V1318 CygS & 20 20 30.5 & +41 21 27 & 1992 Nov 24 & $-80.5$ &  13 & 3  \\
           & 20 20 30.6 & +41 21 26 & 2002 Sep  9 & $5.8$\tablenotemark{b} &  46.7\tablenotemark{b} & 4 \\
PV Cep     & 20 45 21.5 & +67 46 41 & 1984 Jun 26 & $-3$ & $160$ & 5 \\
           &            &           &             & $3$ & $55$ & 5 \\
V1057 Cyg  & 20 57 06.2 & +44 03 46 & 1984 Jan & $\cdots$\tablenotemark{c} & $\cdots$\tablenotemark{c} & 6 \\
\enddata
\tablenotetext{a}{Coordinates are based on the B1900 equatorial system.}
\tablenotetext{b}{We only provide the strongest component. See \citet{Marvel05} for the full list of maser components.}
\tablenotetext{c}{\citet{Rodriguez87} provided an \water\ maser position, but no maser line parameters.}
\tablerefs{(1) \citealt{Schwartz75}; (2) \citealt{Han98}; (3) \citealt{Palla95}; (4) \citealt{Marvel05}
(5) \citealt{Torrelles86}; (6) \citealt{Rodriguez87}; (7) \citealt{Lada81}}
\end{deluxetable}

\clearpage
\begin{deluxetable}{lccrrcccccccc}
\tablecolumns{12}
\tabletypesize{\tiny}
\tablecaption{Observed \water\ and \methanol\ Maser Line Intensities toward BD+40 \degree \label{table:BD40}}
\tablewidth{0pt}
\tablehead{
\colhead{} & 
\colhead{} & 
\colhead{} &
\colhead{} &&
\multicolumn{1}{c}{V1318 CygS} && 
\multicolumn{2}{c}{V1685 Cyg} &&
\multicolumn{2}{c}{V1686 Cyg} \\
\cline{6-6} \cline{8-9} \cline{11-12}
\colhead{Observing} & 
\colhead{Line} & 
\colhead{Frequency} &
\colhead{$v_{\rm LSR}$} &&
\colhead{$F_{\rm peak}$} &&
\colhead{$F_{\rm peak}$} &
\colhead{Theoretical Estimation} &&
\colhead{$F_{\rm peak}$} &
\colhead{Theoretical Estimation}\\
\colhead{Date} & 
\colhead{Number\tablenotemark{a}} & 
\colhead{(\ghz)} & 
\colhead{(\kms)} &&
\colhead{(Jy)} && 
\colhead{(Jy)} & 
\colhead{(Jy)} && 
\colhead{(Jy)} & 
\colhead{(Jy)}
}
\startdata
2010 Apr 18 &      1       & 22.23508  & $-2.11$         && 57.38           && 47.28             & $46.39\pm2.74$ && 55.38           & $55.65\pm1.25$\\
            &      2       & 22.23508  & $-0.21$         && 96.23           && 75.70             & $77.80\pm4.59$ && 91.35           & $93.33\pm2.09$\\
            &      3       & 22.23508  & $+2.74$         && 35.74           && 29.19             & $28.90\pm1.71$ && 35.18           & $34.66\pm0.78$\\
            &      4       & 22.23508  & $+11.17$        &&  5.99           &&  4.77             & $4.84\pm0.29$  &&  5.32           & $5.81\pm0.13$ \\
            &      5       & 44.06943  & $+8.93$         && 54.17           && 26.21             & $23.15\pm5.47$ && 49.99           & $47.93\pm4.30$\\
\hline                                                                                                                    
2010 Oct 20 &      1       & 22.23508  & $-2.42$         && 65.26           && 55.05             & $52.76\pm3.12$ && 63.38           & $63.29\pm1.42$\\
            &      2       & 22.23508  & $-0.11$         && 52.39           && 44.62             & $42.36\pm2.50$ && 49.50           & $50.81\pm1.14$\\
            &      3       & 22.23508  & $+3.48$         && 10.21           &&  8.54             & $8.26\pm0.49$  &&  9.99           & $9.90\pm0.22$\\
\enddata
\tablenotetext{a}{Numbers of each line are displayed in Figure \ref{fig:BD40}.}
\end{deluxetable}

\clearpage
\begin{deluxetable}{lcrcccccccc}
\tablecolumns{10}
\tabletypesize{\scriptsize}
\tablecaption{Observed \water\ Maser Line Intensities toward NGC 7129 Region \label{table:NGC7129}}
\tablewidth{0pt}
\tablehead{
\colhead{} & 
\colhead{} & 
\colhead{} &&
\multicolumn{1}{c}{NGC 7129 FIRS2} && 
\multicolumn{1}{c}{V373 Cep} &&
\multicolumn{2}{c}{V361 Cep} \\
\cline{5-5} \cline{7-7} \cline{9-10}
\colhead{Observing} & 
\colhead{Line} & 
\colhead{$v_{\rm LSR}$} &&
\colhead{$F_{\rm peak}$} &&
\colhead{$F_{\rm peak}$} &&
\colhead{$F_{\rm peak}$} &
\colhead{Theoretical Estimation}\\
\colhead{Date} & 
\colhead{Number\tablenotemark{a}} & 
\colhead{(\kms)} &&
\colhead{(Jy)} && 
\colhead{(Jy)} && 
\colhead{(Jy)} & 
\colhead{(Jy)}
}
\startdata
2010 May 8  &      1       & $-21.80$        && $<$1.33  &&  3.44    && $<$2.33        & $0.62\pm0.10$\\
            &      2       & $-18.43$        && $<$1.33  && 10.66    && 4.55           & $1.92\pm0.32$\\
            &      3       & $-16.54$        && $<$1.33  && 12.54    && 2.44           & $2.26\pm0.38$\\
            &      4       & $-15.06$        && $<$1.33  &&  9.32    && $<$2.33        & $1.68\pm0.28$\\
            &      5       & $-12.53$        && $<$1.33  && 55.28    && 9.99           & $9.96\pm1.67$\\
            &      6       & $-11.27$        && 188.48   && $<$2.55  &&11.43           & $7.31\pm0.30$\\
            &      7       &  $-9.59$        && $<$1.33  && 50.84    && 9.43           & $9.16\pm1.54$\\
            &      8       &  $-7.48$        &&  87.13   && $<$2.55  && 5.11           & $3.38\pm0.14$\\
\enddata
\tablenotetext{a}{Numbers of each line are displayed in Figure \ref{fig:NGC7129}.}
\end{deluxetable}

\clearpage
\begin{deluxetable}{lrccccc}
\tablecolumns{7}
\tabletypesize{\tiny}
\tablecaption{Systemic Velocities and Maser Luminosities of Detected Sources.\label{table:parameter}}
\tablewidth{0pt}
\tablehead{
\colhead{} & 
\colhead{$v_{\rm{sys}}$} &
\colhead{} & 
\colhead{} & 
\colhead{$L_{\water}$\tablenotemark{a}} &
\colhead{$L_{\methanol}$\tablenotemark{a}} \\
\colhead{Source} & 
\colhead{(\kms)} &
\colhead{Transition} & 
\colhead{Ref.} & 
\colhead{($L_{\odot}$)} &
\colhead{($L_{\odot}$)}
}
\startdata
CB 3 & $-$38.8 & ${\rm N_{2}H^{+}}~(J=1-0)$ & 1 &  $[1.5, 3.5, 1.3]\times10^{-6}$& $[7.9, 8.2, 8.3]\times10^{-6}$\\
IRAS 00338+6312 & $-$17.7 & ${\rm CS}~(J=2-1)$ & 2 &$[4.6, 2.6, \cdots]\times10^{-8}$&$[1.1, 1.3, 1.5]\times10^{-7}$ \\
OMC3 MMS9 & +11.6 & ${\rm N_{2}H^{+}}~(J=1-0)$ & 3 & $\cdots$& $[2.2, \cdots, \cdots]\times10^{-8}$\\
OMC2 FIR4 & +11.3 & ${\rm N_{2}H^{+}}~(J=1-0)$ & 3 & $\cdots$ & $[8.2, 9.3, 9.5]\times10^{-7}$\\
IRAS 05338$-$0624 & +7.1 & ${\rm NH_{3}}~(1,1)$ & 4 & $[2.7, 2.9, 2.3]\times10^{-7}$&$[3.0, \cdots, 4.9]\times10^{-8}$\\
IRAS 05338$-$0647 & +9.1 & ${\rm CS}~(J=3-2)$ & 5 & $[4.7, 6.5, 3.6]\times10^{-7}$&$\cdots$ & \\
IRAS 05375$-$0731 & +5.3 & ${\rm NH_{3}}~(1,1)$ & 4 & $[0.24, 0.61, 5.2]\times10^{-7}$ & $\cdots$\\
IRAS 06571$-$0441 & +25.9 & ${\rm CS}~(J=2-1)$  & 2 & $[3.2, \cdots, \cdots]\times10^{-7}$ & $\cdots$ \\
\enddata
\tablenotetext{a}{Numbers in the parentheses represent the maser luminosities in each observation epoch.}
\tablerefs{(1) \citealt{Vries02}; (2) \citealt{Bronfman96}; (3) \citealt{Tatematsu08}; (4) \citealt{Gregorio06}; 
(5) \citealt{Viti06}; (6)\citealt{Ungerechts84}; (7) \citealt{Dent92}; (8) \citealt{Palla95}; (9) \citealt{Scappini96}; (10) \citealt{Yang02}; 
(11)\citealt{Williams99}.}
\end{deluxetable}

\end{document}